
\magnification 1200
\hsize=6.0 true in
\hoffset=0.0 true in
\baselineskip 15pt
\def\lsim{\ ^<\llap{$_\sim$}\ }
\def\gsim{\ ^>\llap{$_\sim$}\ }
\def\w#1{\tilde\omega_#1}
\def\wi{\tilde\omega_i}
\def\x#1{\tilde\chi_#1}
\def\xj{\tilde\chi_j}
\def\tanb{\tan\beta}
\def\sq{\tilde q}
\def\g{\tilde g}
\def\m#1{\tilde {m_#1}}
\def\mH{m_H}
\def\mw{\tilde m_{\omega}}
\def\mx{\tilde m_{\chi}}

\def\Msq{\tilde M_q}
\line{\hfill IFM 8/93}
\line{\hfill TKU-HEP 93/05}
\vskip 1.0 true cm
\centerline{\bf
{\it CP} VIOLATION INDUCED BY SUPERSYMMETRY}
\centerline{\bf
IN $e^+e^-$ COLLISIONS
\footnote{$^*$}{\rm To appear in the proceedings
of the Workshop on Physics and Experiments with
Linear $e^+e^-$ Colliders.}
}
\vskip 1.0 true cm
\centerline{Yoshiki Kizukuri}
\centerline{\it Department of Physics, Tokai University}
\centerline{\it 1117 Kita-Kaname, Hiratsuka 259-12}
\centerline{\it Japan}
\vskip 0.5 true cm
\centerline{and}
\vskip 0.5 true cm
\centerline{Noriyuki Oshimo}
\centerline{\it Grupo Te\'orico de Altas Energias, CFMC}
\centerline{\it Av. Prof. Gama Pinto, 2, 1699 Lisboa Codex}
\centerline{\it Portugal}
\vskip 3.0 true cm
\centerline{\bf Abstract}
\vskip 0.5 true cm
     We discuss the possibility of observing $CP$ violation
predicted by the supersymmetric standard model.
\vfill\eject

     The supersymmetric standard model (SSM) can be
examined from various phenomenological aspects.
Apart from the production and decay of the supersymmetric
particles, this model could make sizable effects on
processes such as the electric dipole moment
(EDM) of the neutron$^{1, 2}$ and the radiative decay of the
$B$-meson$^3$.  It would be useful and necessary to
study the phenomenology in $e^+e^-$ collisions in the
light of the implications of those processes
for which precise measurements have been achieved.
In this context we discuss $CP$ violation.

     In the SSM, $CP$ invariance is generally violated
through the interactions of the supersymmetric particles
as well as the ordinary mechanisms in the standard
model.  These new sources
for $CP$ violation particularly affect the EDMs of
quarks and leptons.   They can be induced already
at one-loop level, whereas in the standard
model the quark EDM receives nonvanishing contributions
at three-loop level.  If the imaginary phases of the
parameters in the lagrangian are not suppressed
and the supersymmetric particles have masses
smaller than O(100) GeV, the EDM of the neutron is
typically predicted$^1$ as $|d_n|\sim 10^{-22}-10^{-23} e$cm.
On the other hand, the experimental bounds$^4$ are
$|d_n|\lsim 10^{-25} e$cm, smaller than the theoretical
value by more than two order of magnitude.
The EDM of the electron is also predicted not to lie
within the experimental bounds$^5$.

     For accomodating these discrepancies
two possible scenarios are available:
(i) The values of the
supersymmetric parameters are nearly real.  (ii) The masses
of the supersymmetric particles are heavy enough.
However, both scenarios
contain problems.  In (i) the Yukawa coupling
constants which give masses to the quarks have to
be complex in order to incorporate $CP$ violation in the
$K$-meson system.  There should exist
some symmetry which, keeping the
Yukawa coupling constants complex, suppresses
the imaginary phases of the other parameters.
In (ii) the supersymmetric particles cannot have
masses much larger than O(1) TeV if the gauge
hierarchy problem should be solved by supersymmetry.
Usually scenario (i) is assumed, though no plausible
symmetry has been proposed.

     We have recently investigated$^2$
the possiblitity of scenario (ii) in detail.
The EDM of the quark receives contributions from the
one-loop diagrams with the squarks $\sq$ and
the charginos $\wi$, the neutralinos $\xj$, or
the gluinos $\g$.  Assuming grand unification and
$N=1$ supergravity, the chargino contribution
generally dominates over the other contributions.
Then, the $CP$-violating effect is parametrized by
the complex phase of the mass parameter,
$\mH=|\mH|{\rm e}^{i\theta}$, which appears in the
bilinear term of Higgs superfields in the
lagrangian.  For $\theta\sim 1$ the EDM of the
neutron have a value within the experimental bounds if
$\mw, \mx\gsim$100 GeV and $\Msq\gsim$1 TeV, $\mw, \mx$,
and $\Msq$ being the masses of the charginos, neutralinos,
and squarks.  The EDM of the electron gives similar
constraints on the masses of the charginos, neutralinos,
and sleptons.  Although the supersymmetric
particles are rather heavy, they could be still
within the mass ranges required for
solving the gauge hierarchy
problem.  The discrepancies concerning the EDMs of the
neutron and the electron may be explained by
scenario (ii) which does not necessitate the
unnatural assumption on the phases of the parameters.

    How the phenomenology in
high energy collider experiments is affected if
the supersymmetric parameters have complex values
and the supersymmetric particles have masses consistent
with the EDMs of the neutron and the electron?
Finding the
squarks or sleptons in near future experiments
would be quite difficult.
Supersymmetry could only be examined directly
by searching for the charginos and/or the
neutralinos in $e^+e^-$ experiments.  The signatures
of these particles have already been studied
extensively in various articles$^6$
under the assumption of real
values for the parameters.  Taking complex
values for these parameters does not change much
the results of these works.  However, the complex
parameters induce $CP$-violating phenomena which
do not occur for the real parameters.

     One of the $CP$-violating effects is
$T$-odd asymmetry$^{7-9}$ in the production and decay
of the charginos or the neutralinos.  Let us
consider the process
$$
         e^+e^-\rightarrow \ell^+\ell^- +X,
$$
where $X$ represents a set of particles
which does not contain
the charged lepton $\ell^+$ nor $\ell^-$.  If the
final state interaction can be neglected, $T$ violation
is examined through the $T$-odd asymmetry of the
angular distribution of the final leptons
$$
  A_T = {d\sigma({\bf p_-\cdot (p_1\times p_2)}>0) -
         d\sigma({\bf p_-\cdot (p_1\times p_2)}<0) \over
         d\sigma({\bf p_-\cdot (p_1\times p_2)}>0) +
         d\sigma({\bf p_-\cdot (p_1\times p_2)}<0)},
\eqno(1)
$$
${\bf p_-, p_1}$,and ${\bf p_2}$ being
the momenta of $e^-, \ell^-$, and $\ell^+$, respectively.
Neglecting the contributions of the squarks and sleptons,
the supersymmetric parameters which determine $A_T$
are the ratio of the vacuum expectation values of the Higgs
bosons, $\tanb$, the SU(2) gaugino mass, $\m2$, and $\mH$.

     The $T$-odd asymmetry can be measured
in the process$^9$
$e^+e^- \rightarrow \xj\x1 \rightarrow \ell^+\ell^-\x1 +\x1$,
where $\x1$ is the lightest neutralino which is assumed to
be stable.  As a typical example for the numerical
evaluation we take $\tanb = 2,
\m2 = 200$ GeV, and $\mH = 200 {\rm e}^{i\pi/4}$ GeV,
which leads to the chargino masses 133, 275 GeV and the
neutralino masses 83, 145, 203, 278 GeV.
At $\sqrt{s}=300$ GeV, the $T$-odd asymmetry
becomes $A_T=-7.6\times 10^{-3}$ if the beams are not
polarized and $A_T=-3.8\times 10^{-2}$ if the electron
beam is left-handedly polarized.

     Larger values of $A_T$ are expected for the
production and decay processes of the
two different charginos$^7$
(a) $e^+e^- \rightarrow \w2^+\w1^- \rightarrow
\ell^+\nu\x1 + \ell^-\bar\nu\x1$ and
(b) $e^+e^- \rightarrow \w2^-\w1^+ \rightarrow
\ell^-\bar\nu\x1 + \ell^+\nu\x1$.
(These two processes should be distinguished from
each other to avoid the reduction of the asymmetry,
which could be done by the cut of the lepton energies.)
For simplicity, we only consider the chargino production
$e^+e^-\rightarrow\w2^+\w1^-$ and give
a rough estimate for $A_T$.  In this
process $T$ violation can be measured by the $T$-odd
asymmetry $\tilde A_T$ which is defined analogously
to Eq. (1) in terms of the polarization vector of
$\w2^+$ and the momenta of $e^-$ and $\w2^+$.
Taking the same parameter values as the previous ones,
at $\sqrt s=500$ GeV the $T$-odd asymmetry becomes
$\tilde A_T=1.2\times 10^{-2}$ in the unpolarized
case and $\tilde A_T=7.5\times 10^{-2}$ in the polarized
case.  The cross section of the whole process (a) or (b) is
roughly of O(1) fb.
As long as $A_T$ is not much reduced compared to
$\tilde A_T$, the $T$-odd asymmetry may be
detectable in next $e^+e^-$ linear colliders.
In the pair production of the same
charginos $T$ violation does not occur at
tree level.  However, the EDM and the $Z$-boson dipole moment
of the chargino are generated at one-loop level, which
break $T$ invariance.  These dipole moments give rise to
$\tilde A_T$ of O($10^{-4}$).

     Another $CP$-violating effect is charge asymmetry$^{10}$
of branching ratios, which could occur, for instance,
in the chargino decay.  Let us assume that
the mass of the charged Higgs bosons is not much different
from that of the $W$-bosons and the chargino can kinematically
decay into both $W^\pm\x1$ and $H^\pm\x1$.
These decays proceed via, in addition
to tree diagrams,
one-loop diagrams by the final state interactions
$W^\pm\x1\rightarrow H^\pm\x1$ and
$H^\pm\x1\rightarrow W^\pm\x1$ in which the charginos are
exchanged.  The interference of the tree and the
one-loop amplitudes causes the difference of the
partial decay rates between $\wi^+$ and $\wi^-$ while
maintaining the same value for their total widths,
i.e. $\Gamma (\wi^+\rightarrow W^+\x1)-\Gamma (\wi^-
\rightarrow W^-\x1)=-\{\Gamma (\wi^+\rightarrow H^+\x1)
-\Gamma (\wi^-\rightarrow H^-\x1)\}\neq 0$.
Since it will not be so difficult to
distinguish $W^\pm$ from $H^\pm$
experimentally, the charge asymmetry may be observed
in the pair production and decay of the chargios.
Calculation shows, however, that the asymmetry is at most
of O($10^{-4}$) which would be too small to be
found in the next $e^+e^-$ colliders.

     In this note we have viewed the SSM from $CP$
violation.  If the imaginary phases of the
supersymmetric parameters have their natural
magnitude of O(1), the EDMs of the neutron and the
electron suggest large masses of O(1)
TeV for the squarks and
the sleptons.  On the other hand,
the masses of the charginos and the neutralinos
could be of O(100) GeV, which can be well explored
at the next $e^+e^-$ linear colliders.
If the charginos or neutralinos are produced, $T$-odd
asymmetry could be induced
in the angular distribution of the final leptons.
Measuring the $T$-odd asymmetry would give us an
important information on the basic parameters,
which will be complimentary to the EDMs
of the neutron and the electron.
\vskip 1.0 true cm
{\bf References}
\smallskip
\item{1.}  J. Ellis, S. Ferrara, and D.V. Nanopoulos,
          {\it Phys. Lett.} {\bf B114} (1982) 231;
\item{ }  W. Buchm\"uller and D. Wyler, {\it Phys. Lett.}
          {\bf B121} (1983) 321;
\item{ }  J. Polchinski and M.B. Wise, {\it Phys. Lett.}
          {\bf B125} (1983) 393;
\item{ }  F. del \'Aguila, M.B. Gavela, J.A. Grifols, and
          A. M\'endez, {\it Phys. Lett.} {\bf B126} (1983) 71.
\item{2.}  Y. Kizukuri and N. Oshimo,
          {\it Phys. Rev.} {\bf D45} (1992) 1806;
          {\bf D46} (1992) 3025.
\item{3.}  N. Oshimo, IFM 12/92
          (to appear in {\it Nucl. Phys.} {\bf B}).
\item{4.}  I.S. Altarev et al., {\it JETP Lett.} {\bf 44} (1986)
          460;
\item{ }  K.F. Smith et al., {\it Phys. Lett.} {\bf B234} (1990)
          191.
\item{5.}  S.A. Murthy et al., {\it Phys. Rev. Lett.} {\bf 63}
          (1989) 965;
\item{ }  D. Cho et al., {\it Phys. Rev. Lett.} {\bf 63}
          (1989) 2559;
\item{ }  K. Abdullah et al., {\it Phys. Rev. Lett.} {\bf 65}
          (1990) 2347.
\item{6.} See e.g. A. Bartl, W. Majerotto, and B. M\"osslacher,
          in {\it Proceedings of the Workshop '$e^+e^-$ Collisions
          at 500 GeV:  The Physics Potential'}, ed. P.M. Zerwas,
          DESY 92-123B (1992).
\item{7.}  Y. Kizukuri, {\it Phys. Lett.} {\bf B193} (1987) 339.
\item{8.}  N. Oshimo, {\it Z. Phys.} {\bf C41} (1988) 129;
          {\it Mod. Phys. Lett.} {\bf A4} (1989) 145.
\item{9.}  Y. Kizukuri and N. Oshimo, {\it Phys. Lett.} {\bf B249}
          (1990) 449;
          in {\it Proceedings of the Workshop '$e^+e^-$ Collisions
          at 500 GeV:  The Physics Potential'}, ed. P.M. Zerwas,
          DESY 92-123B (1992).
\item{10.} N. Oshimo, {\it Phys. Lett.} {\bf B227} (1989) 124.
\vfill\eject
\end